\documentclass[twocolumn, superscriptaddress, amsmath, amssymb]{revtex4-1}

\usepackage{graphicx}
\usepackage{dcolumn}
\usepackage{bm}
\usepackage[T1]{fontenc}       

\usepackage{siunitx}

\begin{document}

\title[ReS2]{Direct observation of the band gap transition in atomically thin ReS$_2$}

\author{Mathias Gehlmann}
\affiliation{PGI-6, Forschungszentrum J\"ulich GmbH, Germany}

\author{Irene Aguilera}
\affiliation{PGI-1/IAS-1, Forschungszentrum J\"ulich GmbH and JARA, Germany}

\author{Gustav Bihlmayer}
\affiliation{PGI-1/IAS-1, Forschungszentrum J\"ulich GmbH and JARA, Germany}

\author{Slavom\'{i}r Nem\v{s}\'{a}k}
\affiliation{PGI-6, Forschungszentrum J\"ulich GmbH, Germany}

\author{Philipp Nagler}
\affiliation{Institut f\"ur Experimentelle und Angewandte Physik, Universit\"at Regensburg, Germany}

\author{Pika Gospodari\v{c}}
\affiliation{PGI-6, Forschungszentrum J\"ulich GmbH, Germany}

\author{Giovanni Zamborlini}
\affiliation{PGI-6, Forschungszentrum J\"ulich GmbH, Germany}

\author{Markus Eschbach}
\affiliation{PGI-6, Forschungszentrum J\"ulich GmbH, Germany}

\author{Vitaliy Feyer}
\affiliation{PGI-6, Forschungszentrum J\"ulich GmbH, Germany}

\author{Florian Kronast}
\affiliation{Abteilung Materialien f\"ur gr\"une Spintronik, Helmholtz-Zentrum Berlin, Germany}

\author{Ewa M\l{}y\'{n}czak}
\affiliation{PGI-6, Forschungszentrum J\"ulich GmbH, Germany}
\affiliation{Faculty of Physics and Applied Computer Science, AGH University of Science and Technology, Poland}

\author{Tobias Korn}
\affiliation{Institut f\"ur Experimentelle und Angewandte Physik, Universit\"at Regensburg, Germany}

\author{Lukasz Plucinski}
\email[Correspondence and requests should be addressed to: \\ ]{ l.plucinski@fz-juelich.de }
\affiliation{PGI-6, Forschungszentrum J\"ulich GmbH, Germany}

\author{Christian Sch\"uller}
\affiliation{Institut f\"ur Experimentelle und Angewandte Physik, Universit\"at Regensburg, Germany}

\author{Stefan Bl\"ugel}
\affiliation{PGI-1/IAS-1, Forschungszentrum J\"ulich GmbH and JARA, Germany}

\author{Claus M. Schneider}
\affiliation{PGI-6, Forschungszentrum J\"ulich GmbH, Germany}

\begin{abstract}
ReS$_2$ is considered as a promising candidate for novel electronic and sensor applications. 
The low crystal symmetry of the van der Waals compound ReS$_2$ leads to a highly anisotropic optical, vibrational, and transport behavior. 
However, the details of the electronic band structure of this fascinating material are still largely unexplored. 
We present a momentum-resolved study of the electronic structure of monolayer, bilayer, and bulk ReS$_2$ using k-space photoemission microscopy in combination with first-principles calculations.
We demonstrate that the valence electrons in bulk ReS$_2$ are -- contrary to assumptions in recent literature -- significantly delocalized across the van der Waals gap. 
Furthermore, we directly observe the evolution of the valence band dispersion as a function of the number of layers, revealing a significantly increased effective electron mass in single-layer crystals. We also find that only bilayer ReS$_2$ has a direct band gap.
Our results establish bilayer ReS$_2$ as a advantageous building block for two-dimensional devices and van der Waals heterostructures.
\end{abstract}

\maketitle

\thispagestyle{empty}

\section{Introduction.} 

\begin{figure*}
\centering
\includegraphics[width=\linewidth]{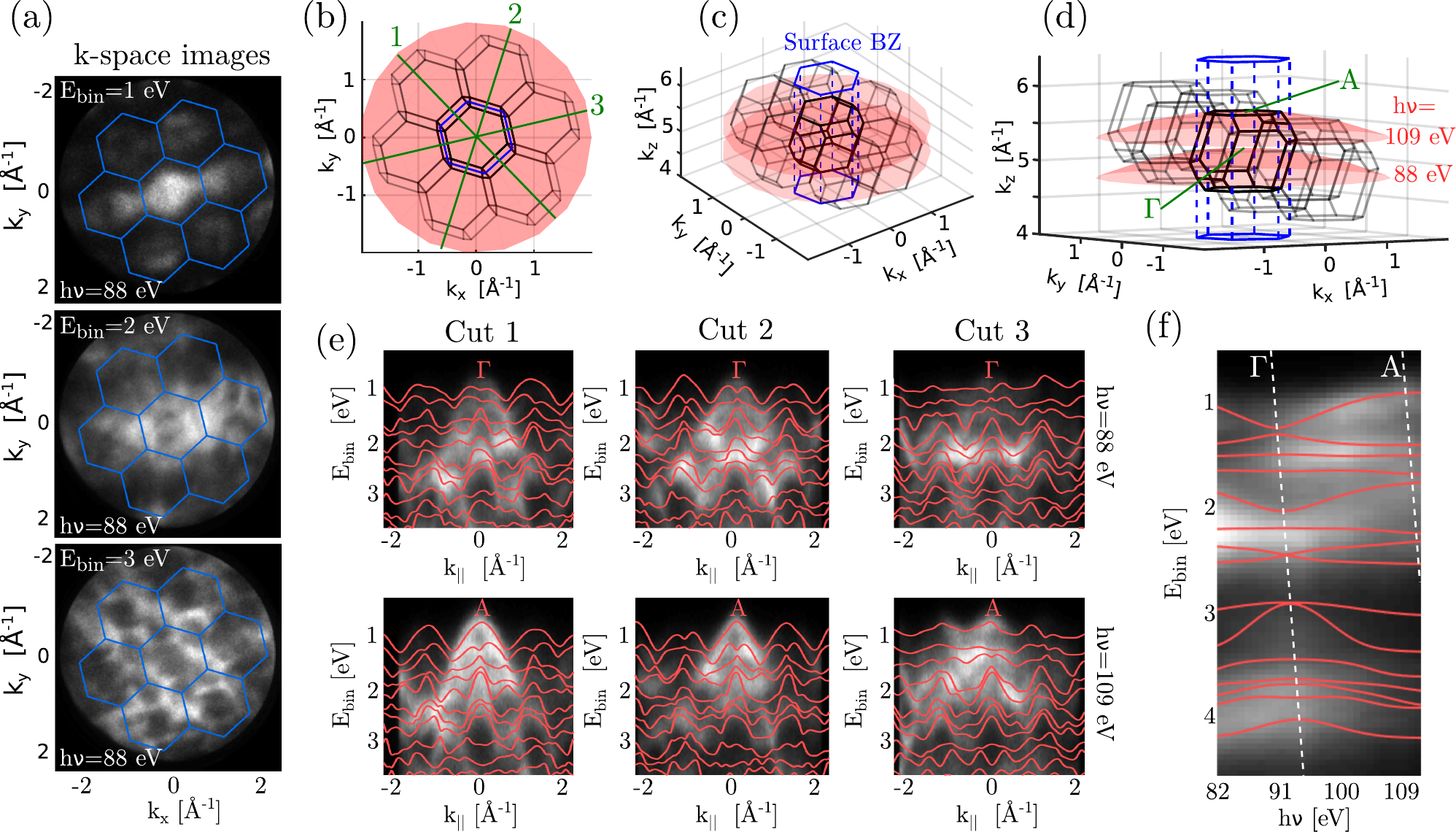}
\caption{
(a) k-space photoemission microscopy images of bulk ReS$_2$ at selected E$_{bin}$ taken with $h\nu = 88~\text{eV}$. The surface BZs are indicated in blue. 
(b-d) Illustration of the origin of the observed photoelectrons within 3D k-space for two selected $h\nu$.
(e) Three selected E$_{bin}$ vs. $k_\parallel$ cuts (see (b)) overlayed with FEFS calculations. At $k_\parallel = 0$ the topmost valence band corresponds to $\mathrm{\Gamma}$/A for $h\nu = 88~\text{eV}~/109~\text{eV}$.
(f) Reconstruction of the $\mathrm{\Gamma}$-A~direction from photoemission spectra taken at different $h\nu = 82 -112~\text{eV}$.
}
\label{fig:3DDelocalization}
\end{figure*}

Transition metal dichalcogenides~(TMDs) hold a tremendous potential for novel electronic and sensoric applications~\cite{PhysTodayVdwPopular, 0953-8984-24-42-424218, doi:10.1038/nature12385}.
Being van der Waals (vdW) compounds they can be thinned down to atomically thin sheets in the same fashion as graphene, but since they include several semiconducting materials with band gaps within the optical range~\cite{doi:10.1021/nl903868w, PhysRevLett.105.136805} they pose a crucial addition to the available ``atomic Legos''~\cite{GeimAtomicScaleLegos} as building blocks for two-dimensional devices and vdW heterostructures~\cite{doi:10.1038/nature12385, doi:10.1021/nl501962c, withers2015lightemittingdiodes, doi:10.1021/nl502075n, Jariwala2017, doi:10.1021/nl504167y, 10.1038/nnano.2010.279, doi:10.1021/nn507278b}.
For several years the majority of research in this field focused on the classical TMDs: MoS$_2$, MoSe$_2$, WS$_2$, and WSe$_2$. 
The bulk crystals of these materials are indirect band gap semiconductors, but undergo a transition into a direct band gap when they are thinned down to the monolayer limit~\cite{wang2012electronicsand, PhysRevLett.105.136805, PhysRevB.84.045409, PhysRevB.88.075409}. 
The direct band gap leads to a drastically increased performance of the monolayer compared to the multilayer equivalents regarding photon absorption and emission, or in transistor devices. 
ReS$_2$ stands out from other TMDs due to its distorted T1 crystal structure\cite{murray1994}, which has a drastically reduced symmetry compared to the 2H structure of classical TMDs.
The low crystal symmetry results in highly anisotropic optical\cite{doi:10.1021/acsphotonics.5b00486}, vibrational\cite{Zhao2015, PSSR:PSSR201510412, doi:10.1021/acs.nanolett.5b04925, doi:10.1021/acsnano.5b07844}, and electric transport properties\cite{doi:10.1021/acsnano.5b04851}, and therefore adds an additional degree of freedom for applications in sensor and electronic devices. 
However, the most basic property for these applications -- the character of the band gap -- has not yet been clarified and received contradicting reports in the recent literature\cite{doi:10.1038/ncomms4252, 2053-1583-3-4-045016, doi:10.1021/acsphotonics.5b00486}. 

The electric transport and optical properties of a material are dominated by the \textit{band edges} of the electronic structure: the valence band maximum~(VBM) and the conduction band minimum~(CBM). 
To our knowledge no experimental studies of the electronic band structure of ReS$_2$ have been published at this point, and the band structure calculations differ in the reported location of the band edges, the evolution of the band dispersion with the number of layers, and in the delocalization of the valence electrons across the vdW gap. 
A recent publication by Tongay~et~al.~\cite{doi:10.1038/ncomms4252} reported a total two-dimensional confinement of the bulk electronic structure, which would make the properties of ReS$_2$ largely independent of its thickness. 
However, this stands in contrast to the observation of strong interlayer vibrational modes in bulk ReS$_2$\cite{C6NR01569G, PSSR:PSSR201510412}, and to reports that the optical absorption\cite{doi:10.1021/acsphotonics.5b00486}, photoluminescence\cite{Zhao2015}, and transport behavior\cite{Ovchinnikov2016} of ultra-thin crystals significantly depend on the number of layers. 

In this article we present a momentum-resolved study of the electronic band structure of monolayer, bilayer, and bulk ReS$_2$. 
Using k-space photoemission microscopy in combination with density functional theory (DFT) calculations we demonstrate a significant three-dimensional delocalization of the valence electrons in bulk ReS$_2$.
We directly observe the layer-dependent evolution of the valence band dispersion in our photoemission experiments, which reveal a relocation of the VBM within the Brillouin zone (BZ), and a drastically increased effective mass of the valence electrons in monolayer ReS$_2$ compared to thicker crystals. 
In addition our many-body $GW$~calculations of the band structure allow for the identification of the CBM, showing that only bilayer ReS$_2$ has a direct band gap. 

\section{Results and discussion.}
To access the three-dimensional valence band dispersion of the ReS$_2$ bulk crystals experimentally we performed k-space microscopy measurements~\cite{Tusche2015520} with different photon energies, and analyzed our data using the free-electron final state (FEFS) model.
The electron momentum components $k_x$ and $k_y$ parallel to the sample surface can be measured directly, since they are conserved in the photoemission process, and the $k_z$~component can be derived from the kinetic energy of the photoelectrons.
Figure~\ref{fig:3DDelocalization}~(a)~shows the k-space images as they are measured by the energy-filtered photoemission electron microscope (PEEM) in momentum mode with the surface BZ indicated in blue. 
One can clearly see that the observed bands do not strictly share the periodicity of the surface BZs. 
On top of intensity variations due to vanishing matrix elements, the position of the bands within each surface BZ is clearly different, since it corresponds to a different position within the bulk BZ in the $k_z$~direction.
This observation is already a strong indication for a delocalization of the valence electrons across the van der Waals gap. 
Figures~\ref{fig:3DDelocalization}~(b-d)~depict the origin of the photoelectrons within three-dimensional k-space according to the FEFS model for two different kinetic energies. 
Since all reciprocal lattice vectors have nonzero z-components, the adjacent bulk BZs are vertically displaced with respect to each other. 
As a result, a constant binding energy $E_{bin}$ cuts through each BZ at a different relative height ($k_z$), which explains the different appearance of all surface BZs in fig.~\ref{fig:3DDelocalization}~(a).

\begin{figure*}
\centering
\includegraphics[width=\linewidth]{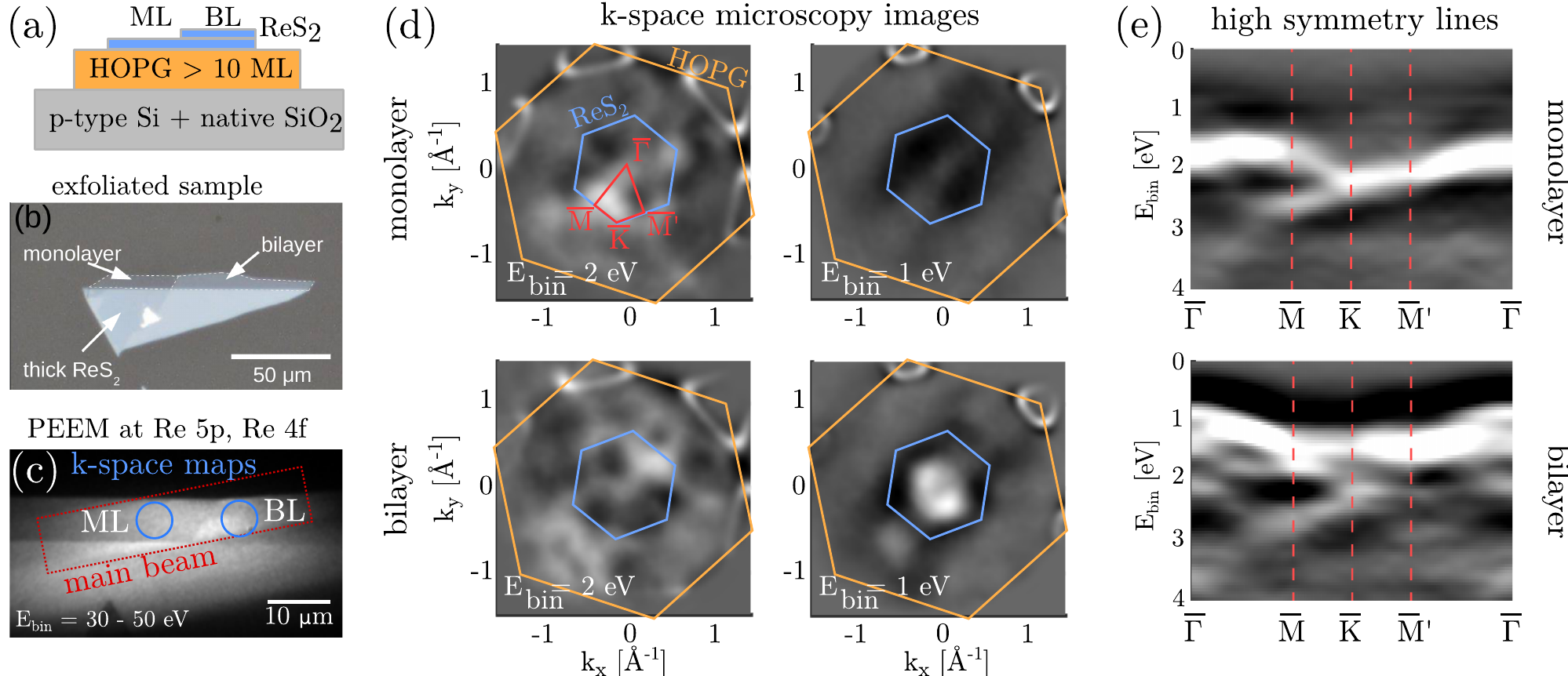}
\caption{
(a) Sketch of the exfoliated few-layer ReS$_2$ sample. 
(b) Optical microscope image of the sample before transfer onto HOPG.
(c) Real space PEEM image with E$_{bin}$ integrated over 30~eV~-~50~eV (Re 5p / Re 4f core levels). 
(d) 2nd derivative k-space images selectively measured on the monolayer and bilayer areas of the sample with the surface BZ of ReS$_2$ indicated in blue and of HOPG in yellow. 
(e) 2nd derivative spectra for monolayer and bilayer along the high symmetry lines as indicated in (d). 
}
\label{fig:ML_BL}
\end{figure*}

For a detailed analysis of the $k_z$ dispersion we repeated the measurement of the valence band spectrum at different photon energies in the range of 82~eV~-~112~eV. 
Figure~\ref{fig:3DDelocalization}~(e) shows three selected $E_{bin}$ versus $k_\parallel$ cuts through the photoemission data~(see fig.~\ref{fig:3DDelocalization}~(b)), each taken with two different photon energies. 
In addition, we performed DFT band structure calculations using the FEFS model to determine $k_z$, and assuming an inner potential $V_0 = 11$~eV, which was obtained by optimizing the overall agreement to the photoemission data. 
For the selected photon energies 88~eV and 109~eV, at $k_\parallel = 0$ the topmost valence band  corresponds to $\mathrm{\Gamma}$ and to the A~point respectively. 
While the photoemission data reveals a pronounced maximum at the A~point, the dispersion at $\mathrm{\Gamma}$ appears rather flat, which agrees well with the calculations.
Figure~\ref{fig:3DDelocalization}~(f)~shows a reconstruction from the photoemission data at $k_\parallel = 0$ from all measured photon energies, and therefore corresponds to the $\mathrm{\Gamma}$-A direction. 
The dispersion of the topmost valence band appears slightly stronger in the calculation than in the photoemission data.
However, the experimentally observed bands in this plot appear rather broad, which is likely caused by the finite sharpness of the $k_z$ selection in the photoemission process. 
Due to the low inelastic mean free path of the photoelectrons, which should be in the order of 2~-~3~\AA ~\cite{SIA:SIA740010103}, one can use the uncertainty principle to estimate $\Delta k_z \approx 0.3~-~0.5$~\AA $^{-1}$ compared to a BZ height of 1.03~\AA $^{-1}$. 
Nevertheless, the experimental data agrees well with the calculations, and also finds the VBM at the A~point and a local band minimum at $\mathrm{\Gamma}$. 
This observation of a strong out-of-plane dispersion is a direct proof of a significant delocalization of the valence electrons across the vdW gap, and shows that the layers in the bulk are not electronically decoupled. 

The observation of a sizable inter-layer interaction also implies a thickness-dependence of the electronic structure. 
We investigated the layer-dependence of the valence band structure by performing further k-space microscopy experiments on exfoliated ReS$_2$ monolayers and bilayers placed onto highly oriented pyrolytic graphite~(HOPG). 
The results are shown in fig.~\ref{fig:ML_BL}. 
The samples were prepared ex-situ and were cleaned for the photoemission experiments purely by degassing at 150$^\circ$C. Samples that were placed directly on SiO$_2$ could not be sufficiently cleaned by this method. 
HOPG is metallic, which suggests that the ReS$_2$ is not sufficiently decoupled from the substrate to be interpreted as freestanding layers. 
However, as one can see in fig.~\ref{fig:ML_BL}~(d) the surface BZ of HOPG is significantly larger than the one of ReS$_2$, and the only HOPG states close to the Fermi level are the s-p-bands at the BZ edges. 
Therefore, the first BZ of ReS$_2$ is unobscured by the HOPG bands, and without matching states for hybridization they can be interpreted as the bands of freestanding ReS$_2$. 
For the bilayer our data reveals a pronounced VBM at the $\overline{\mathrm{\Gamma}}$~point, while the dispersion in the BZ center of the monolayer appears quite flat. 
The drastic differences between the electronic structure of bulk, bilayer, and monolayer ReS$_2$ close to the VBM can be interpreted as a direct result of the increasing quantum confinement in two dimensions as the crystal is thinned down. 
In the monolayer limit the strong dispersion in the $\mathrm{\Gamma}$-A~direction that we found in the bulk crystal is quenched completely, leading to an increased binding energy of the VBM.
This mechanism explains the reports of a blue shift of the optical transitions with a decreasing number of ReS$_2$ layers\cite{doi:10.1038/ncomms4252, doi:10.1021/acsphotonics.5b00486, Zhao2015}, and is the same behavior that was observed for classical TMDs\cite{PhysRevLett.105.136805}, where the VBM is also located at the $\overline{\mathrm{\Gamma}}$~point with a strong $k_z$~dispersion. 
However, it has been unclear if ReS$_2$ also undergoes a band gap transition similar to materials such as MoS$_2$. 

\begin{figure*}
\centering
\includegraphics[width=\linewidth]{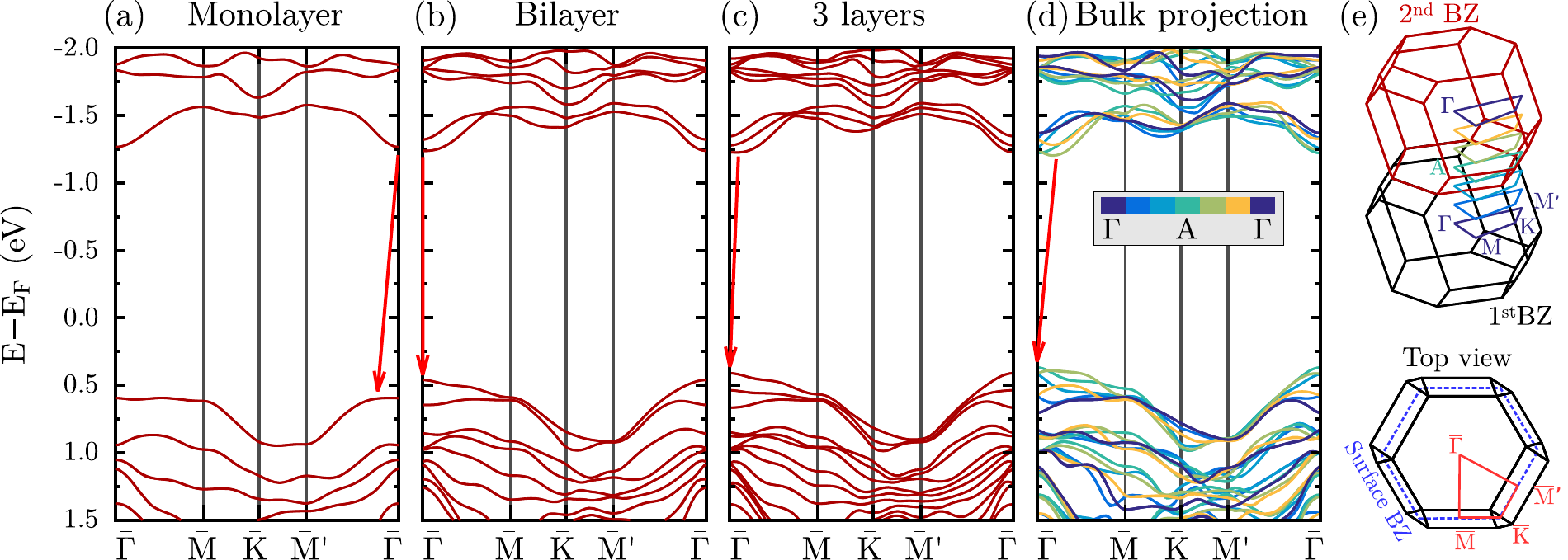}
\caption{$GW$~band structure calculation of monolayer (a), bilayer (b), 3 layers (c), and bulk (d) ReS$_2$ along selected high symmetry directions of the surface BZ. The bulk bands are color coded for the $k_z$~component. The position of the selected directions within 3D~k-space is shown in (e).}
\label{fig:DFT_ML-BL-Bulk}
\end{figure*}

Since our photoemission data only reveals the valence band structure, we have to rely partially on band structure calculations to investigate the band gap character. 
For an accurate treatment of the conduction band structure we performed $GW$~calculations. 
Figure~\ref{fig:DFT_ML-BL-Bulk} shows our $GW$~calculations for a monolayer, a bilayer, and a 3-layer slab, as well as for bulk ReS$_2$ in selected high symmetry directions of the two-dimensional surface BZ. 
The bulk band structure is color coded for the out-of-plane momentum component $k_z$. 
The calculations are in good agreement with the photoemission data, and find the same character of the VBM, which is a distinct maximum at the $\overline{\mathrm{\Gamma}}$~point of the bilayer and a plateau-like region without significant dispersion close to the $\overline{\mathrm{\Gamma}}$~point of the monolayer. 
Based on the CBM that can be identified in our calculations we find an indirect gap for any number of layers except for the bilayer, which has a direct band gap. 
In all cases the band edges are almost degenerate to the smallest direct inter-band transition and are hardly separated in momentum space. 
In classical TMDs the band gap transition is so pronounced because the quantum confinement shifts the bulk VBM at the $\overline{\mathrm{\Gamma}}$~point below the distinct valleys at the BZ edges, which do not exist in ReS$_2$ in this form. 
The indirect character of the gap is by far less pronounced in ReS$_2$, which is a likely explanation why recent studies found indications for both an indirect and a direct gap.
The fundamental band gaps predicted by the $GW$~approximation are 1.85, 1.70, and 1.57~eV for the monolayer, bilayer, and bulk, respectively. 
In terms of the valence band structure our $GW$~calculations are equivalent to our results from DFT. 
It is worth mentioning that the indirect character of the bulk band
structure found in the GW calculations is a result of the many-body
effects that are not incorporated in the GGA calculations. The
conduction band structure in the GGA calculations is slightly different
from GW and shows a direct band gap at the A-point (see supporting information for comparison).

Another interesting aspect that is revealed by our results is that the thinning step from bilayer to monolayer, which enforces the complete confinement of the valence electrons within two dimensions, is accompanied by a significant decrease of the valence band dispersion in the x-y-directions as well. 
This observation can be interpreted as an increased effective mass of the electrons close to the VBM. 
Recently, it was found in layer-dependent transport measurements that monolayers of ReS$_2$ have a drastically reduced conductivity compared to bilayers or thicker crystals\cite{Ovchinnikov2016}. 
Apart from the larger band gap, this behavior was attributed to a significantly stronger influence of impurities and the interface. 
Our results rather suggest an intrinsic mechanism that suppresses the delocalization of VBM electrons within the x-y-plane.
This behavior could significantly impact the electron mobility for single monolayer crystals.

\section{Conclusions.}
Using a combination of k-space photoemission microscopy and band structure calculations we have presented a thorough, momentum-resolved study of the electronic band structure of monolayer, bilayer, and bulk ReS$_2$. 
By showing a significant out-of-plane delocalization of the valence electrons we have demonstrated that --~contrary to assumptions in the recent literature~-- the layers of bulk ReS$_2$ are not electronically decoupled.
We identified the layer-dependent position of the VBM in our photoemission data, and in combination with our band structure calculations we concluded that only bilayer ReS$_2$ has a direct band gap, although the indirect character of the gap for other crystal thicknesses is far less pronounced than in classical TMDs. 
Furthermore, our results have shown that in monolayer ReS$_2$ the valence electrons at the VBM suffer from a drastic increase of their effective mass, suggesting an intrinsic mechanism that could reduce the electron mobility in the in-plane directions compared to bilayers or thicker crystals. 
For many applications this feature in combination with the direct band gap would make ReS$_2$ bilayers a preferred choice in comparison to monolayers or thicker crystals.

\section{Methods}
\subsection{Sample Preparation.} The ReS$_2$ single crystal was bought from \textit{HQGraphene} and cleaved in air with adhesive tape immediately before it was introduced into ultra-high vacuum and degassed at 250$^\circ$C. 
Atomically thin ReS$_2$ samples were produced ex situ by mechanical exfoliation from a bulk crystal. 
Using an all dry viscoelastic stamp method, which is described in Ref.~\cite{2053-1583-1-1-011002}, a ReS$_2$ sample that simultaneously contained monolayer as well as bilayer areas was transfered onto HOPG. 
The thickness of the ultra-thin sample areas was identified by the optical contrast in the transmission mode of an optical microscope.
In situ the atomically thin samples were degassed for one hour at 150$^\circ$C.
PEEM images of the samples and optical microscope images taken before and after the transfer process are shown in the supporting information.

\subsection{Photoemission Experiments.} The photoemission experiments on the ReS$_2$ bulk crystal were performed at the \textit{NanoESCA beamline} of Elettra, Trieste, using a modified \textit{FOCUS NanoESCA}. 
Homogeneous areas on the cleaved surface were identified using PEEM in real space mode and Hg lamp as excitation source.
The lateral resolution of the k-resolved photoemission maps was limited by the beam size ($\SI{10}{\micro\meter} \times \SI{10}{\micro\meter}$) ruling out the influence of step edges and grain boundaries.
The total energy resolution was better than~100~meV.
Due to the highly focused synchrotron beam on the sample, the spectra suffered from charging effects, which were corrected in the data by an energy offset of $100 - 150$~meV.
The photoemission experiments on the atomically thin ReS$_2$ samples were performed at the \textit{UE49 SPEEM} endstation of \textit{BESSY~II}, Berlin. 
Using the field aperture of the PEEM the field of view in real space was limited to $\approx$~\SI{5}{\micro\meter}. 
This way the monolayer and bilayer parts of the sample could be selectively measured in momentum mode. 
The total energy resolution was~300~meV. 

\subsection{Band structure calculations.} 
Calculations were carried out with the DFT code {\sc fleur}{}~\cite{FLEURhomepage} and the $GW$ code {\sc spex}{}~\cite{friedrich2010-spex}, which use the all-electron full-potential linearized augmented-plane-wave (FLAPW) formalism. 
The electron density was obtained with the Perdew-Burke-Ernzerhof (PBE) functional. We used the experimental lattice structure of Ref.~\cite{murray1994}. 
The calculations of the monolayer, bilayer, and 3-layer slab were performed with a tight-binding Hamiltonian with \textit{ab-initio} parameters obtained from the bulk $GW$ calculation. 
Maximally localized Wannier functions, generated by the {\sc wannier90}{} library,\cite{wannier90} were used as a basis.


\section{Acknowledgements}
The authors thank the Helmholtz-Zentrum Berlin and Elettra Sincrotrone Trieste for providing synchrotron radiation. 
The financial support by the Helmholtz Association via the \textit{Initiative and Networking Fund} and via the \textit{Virtual Institute for Topological Insulators}, and by the DFG via GRK1570 is gratefully acknowledged.

\bibliography{bib_mathias_iso}

\begin{thebibliography}{33}%
\makeatletter
\providecommand \@ifxundefined [1]{%
 \@ifx{#1\undefined}
}%
\providecommand \@ifnum [1]{%
 \ifnum #1\expandafter \@firstoftwo
 \else \expandafter \@secondoftwo
 \fi
}%
\providecommand \@ifx [1]{%
 \ifx #1\expandafter \@firstoftwo
 \else \expandafter \@secondoftwo
 \fi
}%
\providecommand \natexlab [1]{#1}%
\providecommand \enquote  [1]{``#1''}%
\providecommand \bibnamefont  [1]{#1}%
\providecommand \bibfnamefont [1]{#1}%
\providecommand \citenamefont [1]{#1}%
\providecommand \href@noop [0]{\@secondoftwo}%
\providecommand \href [0]{\begingroup \@sanitize@url \@href}%
\providecommand \@href[1]{\@@startlink{#1}\@@href}%
\providecommand \@@href[1]{\endgroup#1\@@endlink}%
\providecommand \@sanitize@url [0]{\catcode `\\12\catcode `\$12\catcode
  `\&12\catcode `\#12\catcode `\^12\catcode `\_12\catcode `\%12\relax}%
\providecommand \@@startlink[1]{}%
\providecommand \@@endlink[0]{}%
\providecommand \url  [0]{\begingroup\@sanitize@url \@url }%
\providecommand \@url [1]{\endgroup\@href {#1}{\urlprefix }}%
\providecommand \urlprefix  [0]{URL }%
\providecommand \Eprint [0]{\href }%
\providecommand \doibase [0]{http://dx.doi.org/}%
\providecommand \selectlanguage [0]{\@gobble}%
\providecommand \bibinfo  [0]{\@secondoftwo}%
\providecommand \bibfield  [0]{\@secondoftwo}%
\providecommand \translation [1]{[#1]}%
\providecommand \BibitemOpen [0]{}%
\providecommand \bibitemStop [0]{}%
\providecommand \bibitemNoStop [0]{.\EOS\space}%
\providecommand \EOS [0]{\spacefactor3000\relax}%
\providecommand \BibitemShut  [1]{\csname bibitem#1\endcsname}%
\let\auto@bib@innerbib\@empty
\bibitem [{\citenamefont {Ajayan}\ \emph {et~al.}(2016)\citenamefont {Ajayan},
  \citenamefont {Kim},\ and\ \citenamefont {Banerjee}}]{PhysTodayVdwPopular}%
  \BibitemOpen
  \bibfield  {author} {\bibinfo {author} {\bibfnamefont {P.}~\bibnamefont
  {Ajayan}}, \bibinfo {author} {\bibfnamefont {P.}~\bibnamefont {Kim}}, \ and\
  \bibinfo {author} {\bibfnamefont {K.}~\bibnamefont {Banerjee}},\ }\href
  {\doibase 10.1063/PT.3.3297} {\bibfield  {journal} {\bibinfo  {journal}
  {Phys. Today}\ }\textbf {\bibinfo {volume} {69}},\ \bibinfo {pages} {38}
  (\bibinfo {year} {2016})}\BibitemShut {NoStop}%
\bibitem [{\citenamefont {Björkman}\ \emph {et~al.}(2012)\citenamefont
  {Björkman}, \citenamefont {Gulans}, \citenamefont {Krasheninnikov},\ and\
  \citenamefont {Nieminen}}]{0953-8984-24-42-424218}%
  \BibitemOpen
  \bibfield  {author} {\bibinfo {author} {\bibfnamefont {T.}~\bibnamefont
  {Björkman}}, \bibinfo {author} {\bibfnamefont {A.}~\bibnamefont {Gulans}},
  \bibinfo {author} {\bibfnamefont {A.~V.}\ \bibnamefont {Krasheninnikov}}, \
  and\ \bibinfo {author} {\bibfnamefont {R.~M.}\ \bibnamefont {Nieminen}},\
  }\href {http://stacks.iop.org/0953-8984/24/i=42/a=424218} {\bibfield
  {journal} {\bibinfo  {journal} {Journal of Physics: Condensed Matter}\
  }\textbf {\bibinfo {volume} {24}},\ \bibinfo {pages} {424218} (\bibinfo
  {year} {2012})}\BibitemShut {NoStop}%
\bibitem [{\citenamefont {Geim}\ and\ \citenamefont
  {Grigorieva}(2013)}]{doi:10.1038/nature12385}%
  \BibitemOpen
  \bibfield  {author} {\bibinfo {author} {\bibfnamefont {A.~K.}\ \bibnamefont
  {Geim}}\ and\ \bibinfo {author} {\bibfnamefont {I.~V.}\ \bibnamefont
  {Grigorieva}},\ }\href {\doibase 10.1038/nature12385} {\bibfield  {journal}
  {\bibinfo  {journal} {Nature}\ }\textbf {\bibinfo {volume} {499}},\ \bibinfo
  {pages} {419} (\bibinfo {year} {2013})}\BibitemShut {NoStop}%
\bibitem [{\citenamefont {Splendiani}\ \emph {et~al.}(2010)\citenamefont
  {Splendiani}, \citenamefont {Sun}, \citenamefont {Zhang}, \citenamefont {Li},
  \citenamefont {Kim}, \citenamefont {Chim}, \citenamefont {Galli},\ and\
  \citenamefont {Wang}}]{doi:10.1021/nl903868w}%
  \BibitemOpen
  \bibfield  {author} {\bibinfo {author} {\bibfnamefont {A.}~\bibnamefont
  {Splendiani}}, \bibinfo {author} {\bibfnamefont {L.}~\bibnamefont {Sun}},
  \bibinfo {author} {\bibfnamefont {Y.}~\bibnamefont {Zhang}}, \bibinfo
  {author} {\bibfnamefont {T.}~\bibnamefont {Li}}, \bibinfo {author}
  {\bibfnamefont {J.}~\bibnamefont {Kim}}, \bibinfo {author} {\bibfnamefont
  {C.-Y.}\ \bibnamefont {Chim}}, \bibinfo {author} {\bibfnamefont
  {G.}~\bibnamefont {Galli}}, \ and\ \bibinfo {author} {\bibfnamefont
  {F.}~\bibnamefont {Wang}},\ }\href {\doibase 10.1021/nl903868w} {\bibfield
  {journal} {\bibinfo  {journal} {Nano Letters}\ }\textbf {\bibinfo {volume}
  {10}},\ \bibinfo {pages} {1271} (\bibinfo {year} {2010})},\ \Eprint
  {http://arxiv.org/abs/http://dx.doi.org/10.1021/nl903868w}
  {http://dx.doi.org/10.1021/nl903868w} \BibitemShut {NoStop}%
\bibitem [{\citenamefont {Mak}\ \emph {et~al.}(2010)\citenamefont {Mak},
  \citenamefont {Lee}, \citenamefont {Hone}, \citenamefont {Shan},\ and\
  \citenamefont {Heinz}}]{PhysRevLett.105.136805}%
  \BibitemOpen
  \bibfield  {author} {\bibinfo {author} {\bibfnamefont {K.~F.}\ \bibnamefont
  {Mak}}, \bibinfo {author} {\bibfnamefont {C.}~\bibnamefont {Lee}}, \bibinfo
  {author} {\bibfnamefont {J.}~\bibnamefont {Hone}}, \bibinfo {author}
  {\bibfnamefont {J.}~\bibnamefont {Shan}}, \ and\ \bibinfo {author}
  {\bibfnamefont {T.~F.}\ \bibnamefont {Heinz}},\ }\href {\doibase
  10.1103/PhysRevLett.105.136805} {\bibfield  {journal} {\bibinfo  {journal}
  {Phys. Rev. Lett.}\ }\textbf {\bibinfo {volume} {105}},\ \bibinfo {pages}
  {136805} (\bibinfo {year} {2010})}\BibitemShut {NoStop}%
\bibitem [{\citenamefont {Geim}(2014)}]{GeimAtomicScaleLegos}%
  \BibitemOpen
  \bibfield  {author} {\bibinfo {author} {\bibfnamefont {A.~K.}\ \bibnamefont
  {Geim}},\ }\href@noop {} {\bibfield  {journal} {\bibinfo  {journal} {SciAm}\
  }\textbf {\bibinfo {volume} {311}},\ \bibinfo {pages} {50} (\bibinfo {year}
  {2014})}\BibitemShut {NoStop}%
\bibitem [{\citenamefont {Furchi}\ \emph {et~al.}(2014)\citenamefont {Furchi},
  \citenamefont {Pospischil}, \citenamefont {Libisch}, \citenamefont
  {Burgdörfer},\ and\ \citenamefont {Mueller}}]{doi:10.1021/nl501962c}%
  \BibitemOpen
  \bibfield  {author} {\bibinfo {author} {\bibfnamefont {M.~M.}\ \bibnamefont
  {Furchi}}, \bibinfo {author} {\bibfnamefont {A.}~\bibnamefont {Pospischil}},
  \bibinfo {author} {\bibfnamefont {F.}~\bibnamefont {Libisch}}, \bibinfo
  {author} {\bibfnamefont {J.}~\bibnamefont {Burgdörfer}}, \ and\ \bibinfo
  {author} {\bibfnamefont {T.}~\bibnamefont {Mueller}},\ }\href {\doibase
  10.1021/nl501962c} {\bibfield  {journal} {\bibinfo  {journal} {Nano Letters}\
  }\textbf {\bibinfo {volume} {14}},\ \bibinfo {pages} {4785} (\bibinfo {year}
  {2014})},\ \Eprint {http://arxiv.org/abs/http://dx.doi.org/10.1021/nl501962c}
  {http://dx.doi.org/10.1021/nl501962c} \BibitemShut {NoStop}%
\bibitem [{\citenamefont {Withers}\ \emph {et~al.}(2015)\citenamefont
  {Withers}, \citenamefont {Pozo}, \citenamefont {Mishchenko}, \citenamefont
  {Rooney}, \citenamefont {Gholinia}, \citenamefont {Watanabe}, \citenamefont
  {Taniguchi}, \citenamefont {Haigh}, \citenamefont {Geim}, \citenamefont
  {Tartakovskii},\ and\ \citenamefont
  {Novoselov}}]{withers2015lightemittingdiodes}%
  \BibitemOpen
  \bibfield  {author} {\bibinfo {author} {\bibfnamefont {F.}~\bibnamefont
  {Withers}}, \bibinfo {author} {\bibfnamefont {D.}~\bibnamefont {Pozo}},
  \bibinfo {author} {\bibfnamefont {A.}~\bibnamefont {Mishchenko}}, \bibinfo
  {author} {\bibfnamefont {A.}~\bibnamefont {Rooney}}, \bibinfo {author}
  {\bibfnamefont {A.}~\bibnamefont {Gholinia}}, \bibinfo {author}
  {\bibfnamefont {K.}~\bibnamefont {Watanabe}}, \bibinfo {author}
  {\bibfnamefont {T.}~\bibnamefont {Taniguchi}}, \bibinfo {author}
  {\bibfnamefont {S.}~\bibnamefont {Haigh}}, \bibinfo {author} {\bibfnamefont
  {A.}~\bibnamefont {Geim}}, \bibinfo {author} {\bibfnamefont {A.}~\bibnamefont
  {Tartakovskii}}, \ and\ \bibinfo {author} {\bibfnamefont {K.}~\bibnamefont
  {Novoselov}},\ }\href {\doibase Letter} {\bibfield  {journal} {\bibinfo
  {journal} {Nat. Mater.}\ }\textbf {\bibinfo {volume} {14}},\ \bibinfo {pages}
  {301} (\bibinfo {year} {2015})}\BibitemShut {NoStop}%
\bibitem [{\citenamefont {Cheng}\ \emph {et~al.}(2014)\citenamefont {Cheng},
  \citenamefont {Li}, \citenamefont {Zhou}, \citenamefont {Wang}, \citenamefont
  {Yin}, \citenamefont {Jiang}, \citenamefont {Liu}, \citenamefont {Chen},
  \citenamefont {Huang},\ and\ \citenamefont {Duan}}]{doi:10.1021/nl502075n}%
  \BibitemOpen
  \bibfield  {author} {\bibinfo {author} {\bibfnamefont {R.}~\bibnamefont
  {Cheng}}, \bibinfo {author} {\bibfnamefont {D.}~\bibnamefont {Li}}, \bibinfo
  {author} {\bibfnamefont {H.}~\bibnamefont {Zhou}}, \bibinfo {author}
  {\bibfnamefont {C.}~\bibnamefont {Wang}}, \bibinfo {author} {\bibfnamefont
  {A.}~\bibnamefont {Yin}}, \bibinfo {author} {\bibfnamefont {S.}~\bibnamefont
  {Jiang}}, \bibinfo {author} {\bibfnamefont {Y.}~\bibnamefont {Liu}}, \bibinfo
  {author} {\bibfnamefont {Y.}~\bibnamefont {Chen}}, \bibinfo {author}
  {\bibfnamefont {Y.}~\bibnamefont {Huang}}, \ and\ \bibinfo {author}
  {\bibfnamefont {X.}~\bibnamefont {Duan}},\ }\href {\doibase
  10.1021/nl502075n} {\bibfield  {journal} {\bibinfo  {journal} {Nano Letters}\
  }\textbf {\bibinfo {volume} {14}},\ \bibinfo {pages} {5590} (\bibinfo {year}
  {2014})},\ \Eprint {http://arxiv.org/abs/http://dx.doi.org/10.1021/nl502075n}
  {http://dx.doi.org/10.1021/nl502075n} \BibitemShut {NoStop}%
\bibitem [{\citenamefont {Jariwala}\ \emph {et~al.}(2017)\citenamefont
  {Jariwala}, \citenamefont {Marks},\ and\ \citenamefont
  {Hersam}}]{Jariwala2017}%
  \BibitemOpen
  \bibfield  {author} {\bibinfo {author} {\bibfnamefont {D.}~\bibnamefont
  {Jariwala}}, \bibinfo {author} {\bibfnamefont {T.~J.}\ \bibnamefont {Marks}},
  \ and\ \bibinfo {author} {\bibfnamefont {M.~C.}\ \bibnamefont {Hersam}},\
  }\href {http://dx.doi.org/10.1038/nmat4703} {\bibfield  {journal} {\bibinfo
  {journal} {Nat. Mater.}\ }\textbf {\bibinfo {volume} {16}},\ \bibinfo {pages}
  {170} (\bibinfo {year} {2017})}\BibitemShut {NoStop}%
\bibitem [{\citenamefont {Coy~Diaz}\ \emph {et~al.}(2015)\citenamefont
  {Coy~Diaz}, \citenamefont {Avila}, \citenamefont {Chen}, \citenamefont
  {Addou}, \citenamefont {Asensio},\ and\ \citenamefont
  {Batzill}}]{doi:10.1021/nl504167y}%
  \BibitemOpen
  \bibfield  {author} {\bibinfo {author} {\bibfnamefont {H.}~\bibnamefont
  {Coy~Diaz}}, \bibinfo {author} {\bibfnamefont {J.}~\bibnamefont {Avila}},
  \bibinfo {author} {\bibfnamefont {C.}~\bibnamefont {Chen}}, \bibinfo {author}
  {\bibfnamefont {R.}~\bibnamefont {Addou}}, \bibinfo {author} {\bibfnamefont
  {M.~C.}\ \bibnamefont {Asensio}}, \ and\ \bibinfo {author} {\bibfnamefont
  {M.}~\bibnamefont {Batzill}},\ }\href {\doibase 10.1021/nl504167y} {\bibfield
   {journal} {\bibinfo  {journal} {Nano Letters}\ }\textbf {\bibinfo {volume}
  {15}},\ \bibinfo {pages} {1135} (\bibinfo {year} {2015})}\BibitemShut
  {NoStop}%
\bibitem [{\citenamefont {Radisavljevic}\ \emph {et~al.}(2011)\citenamefont
  {Radisavljevic}, \citenamefont {Radenovic}, \citenamefont {Brivio},
  \citenamefont {Giacometti},\ and\ \citenamefont
  {Kis}}]{10.1038/nnano.2010.279}%
  \BibitemOpen
  \bibfield  {author} {\bibinfo {author} {\bibfnamefont {B.}~\bibnamefont
  {Radisavljevic}}, \bibinfo {author} {\bibfnamefont {A.}~\bibnamefont
  {Radenovic}}, \bibinfo {author} {\bibfnamefont {J.}~\bibnamefont {Brivio}},
  \bibinfo {author} {\bibfnamefont {V.}~\bibnamefont {Giacometti}}, \ and\
  \bibinfo {author} {\bibfnamefont {A.}~\bibnamefont {Kis}},\ }\href@noop {}
  {\bibfield  {journal} {\bibinfo  {journal} {Nat. Nano}\ }\textbf {\bibinfo
  {volume} {6}},\ \bibinfo {pages} {147} (\bibinfo {year} {2011})}\BibitemShut
  {NoStop}%
\bibitem [{\citenamefont {Roy}\ \emph {et~al.}(2015)\citenamefont {Roy},
  \citenamefont {Tosun}, \citenamefont {Cao}, \citenamefont {Fang},
  \citenamefont {Lien}, \citenamefont {Zhao}, \citenamefont {Chen},
  \citenamefont {Chueh}, \citenamefont {Guo},\ and\ \citenamefont
  {Javey}}]{doi:10.1021/nn507278b}%
  \BibitemOpen
  \bibfield  {author} {\bibinfo {author} {\bibfnamefont {T.}~\bibnamefont
  {Roy}}, \bibinfo {author} {\bibfnamefont {M.}~\bibnamefont {Tosun}}, \bibinfo
  {author} {\bibfnamefont {X.}~\bibnamefont {Cao}}, \bibinfo {author}
  {\bibfnamefont {H.}~\bibnamefont {Fang}}, \bibinfo {author} {\bibfnamefont
  {D.-H.}\ \bibnamefont {Lien}}, \bibinfo {author} {\bibfnamefont
  {P.}~\bibnamefont {Zhao}}, \bibinfo {author} {\bibfnamefont {Y.-Z.}\
  \bibnamefont {Chen}}, \bibinfo {author} {\bibfnamefont {Y.-L.}\ \bibnamefont
  {Chueh}}, \bibinfo {author} {\bibfnamefont {J.}~\bibnamefont {Guo}}, \ and\
  \bibinfo {author} {\bibfnamefont {A.}~\bibnamefont {Javey}},\ }\href
  {\doibase 10.1021/nn507278b} {\bibfield  {journal} {\bibinfo  {journal} {ACS
  Nano}\ }\textbf {\bibinfo {volume} {9}},\ \bibinfo {pages} {2071} (\bibinfo
  {year} {2015})},\ \Eprint
  {http://arxiv.org/abs/http://dx.doi.org/10.1021/nn507278b}
  {http://dx.doi.org/10.1021/nn507278b} \BibitemShut {NoStop}%
\bibitem [{\citenamefont {Wang}\ \emph {et~al.}(2012)\citenamefont {Wang},
  \citenamefont {Kalantar}, \citenamefont {Kis}, \citenamefont {Coleman},\ and\
  \citenamefont {Strano}}]{wang2012electronicsand}%
  \BibitemOpen
  \bibfield  {author} {\bibinfo {author} {\bibfnamefont {Q.~H.}\ \bibnamefont
  {Wang}}, \bibinfo {author} {\bibnamefont {Kalantar}}, \bibinfo {author}
  {\bibfnamefont {A.}~\bibnamefont {Kis}}, \bibinfo {author} {\bibfnamefont
  {J.~N.}\ \bibnamefont {Coleman}}, \ and\ \bibinfo {author} {\bibfnamefont
  {M.~S.}\ \bibnamefont {Strano}},\ }\href {\doibase 10.1038/nnano.2012.193}
  {\bibfield  {journal} {\bibinfo  {journal} {Nat. Nano}\ }\textbf {\bibinfo
  {volume} {7}},\ \bibinfo {pages} {699} (\bibinfo {year} {2012})}\BibitemShut
  {NoStop}%
\bibitem [{\citenamefont {Han}\ \emph {et~al.}(2011)\citenamefont {Han},
  \citenamefont {Kwon}, \citenamefont {Kim}, \citenamefont {Ryu}, \citenamefont
  {Yun}, \citenamefont {Kim}, \citenamefont {Hwang}, \citenamefont {Kang},
  \citenamefont {Baik}, \citenamefont {Shin},\ and\ \citenamefont
  {Hong}}]{PhysRevB.84.045409}%
  \BibitemOpen
  \bibfield  {author} {\bibinfo {author} {\bibfnamefont {S.~W.}\ \bibnamefont
  {Han}}, \bibinfo {author} {\bibfnamefont {H.}~\bibnamefont {Kwon}}, \bibinfo
  {author} {\bibfnamefont {S.~K.}\ \bibnamefont {Kim}}, \bibinfo {author}
  {\bibfnamefont {S.}~\bibnamefont {Ryu}}, \bibinfo {author} {\bibfnamefont
  {W.~S.}\ \bibnamefont {Yun}}, \bibinfo {author} {\bibfnamefont {D.~H.}\
  \bibnamefont {Kim}}, \bibinfo {author} {\bibfnamefont {J.~H.}\ \bibnamefont
  {Hwang}}, \bibinfo {author} {\bibfnamefont {J.-S.}\ \bibnamefont {Kang}},
  \bibinfo {author} {\bibfnamefont {J.}~\bibnamefont {Baik}}, \bibinfo {author}
  {\bibfnamefont {H.~J.}\ \bibnamefont {Shin}}, \ and\ \bibinfo {author}
  {\bibfnamefont {S.~C.}\ \bibnamefont {Hong}},\ }\href {\doibase
  10.1103/PhysRevB.84.045409} {\bibfield  {journal} {\bibinfo  {journal} {Phys.
  Rev. B}\ }\textbf {\bibinfo {volume} {84}},\ \bibinfo {pages} {045409}
  (\bibinfo {year} {2011})}\BibitemShut {NoStop}%
\bibitem [{\citenamefont {Cappelluti}\ \emph {et~al.}(2013)\citenamefont
  {Cappelluti}, \citenamefont {Rold\'an}, \citenamefont {Silva-Guill\'en},
  \citenamefont {Ordej\'on},\ and\ \citenamefont
  {Guinea}}]{PhysRevB.88.075409}%
  \BibitemOpen
  \bibfield  {author} {\bibinfo {author} {\bibfnamefont {E.}~\bibnamefont
  {Cappelluti}}, \bibinfo {author} {\bibfnamefont {R.}~\bibnamefont
  {Rold\'an}}, \bibinfo {author} {\bibfnamefont {J.~A.}\ \bibnamefont
  {Silva-Guill\'en}}, \bibinfo {author} {\bibfnamefont {P.}~\bibnamefont
  {Ordej\'on}}, \ and\ \bibinfo {author} {\bibfnamefont {F.}~\bibnamefont
  {Guinea}},\ }\href {\doibase 10.1103/PhysRevB.88.075409} {\bibfield
  {journal} {\bibinfo  {journal} {Phys. Rev. B}\ }\textbf {\bibinfo {volume}
  {88}},\ \bibinfo {pages} {075409} (\bibinfo {year} {2013})}\BibitemShut
  {NoStop}%
\bibitem [{\citenamefont {Murray}\ \emph {et~al.}(1994)\citenamefont {Murray},
  \citenamefont {Kelty}, \citenamefont {Chianelli},\ and\ \citenamefont
  {Day}}]{murray1994}%
  \BibitemOpen
  \bibfield  {author} {\bibinfo {author} {\bibfnamefont {H.}~\bibnamefont
  {Murray}}, \bibinfo {author} {\bibfnamefont {S.}~\bibnamefont {Kelty}},
  \bibinfo {author} {\bibfnamefont {R.}~\bibnamefont {Chianelli}}, \ and\
  \bibinfo {author} {\bibfnamefont {C.}~\bibnamefont {Day}},\ }\href@noop {}
  {\bibfield  {journal} {\bibinfo  {journal} {Inorganic Chemistry}\ }\textbf
  {\bibinfo {volume} {33}},\ \bibinfo {pages} {4418} (\bibinfo {year}
  {1994})}\BibitemShut {NoStop}%
\bibitem [{\citenamefont {Aslan}\ \emph {et~al.}(2016)\citenamefont {Aslan},
  \citenamefont {Chenet}, \citenamefont {van~der Zande}, \citenamefont {Hone},\
  and\ \citenamefont {Heinz}}]{doi:10.1021/acsphotonics.5b00486}%
  \BibitemOpen
  \bibfield  {author} {\bibinfo {author} {\bibfnamefont {O.~B.}\ \bibnamefont
  {Aslan}}, \bibinfo {author} {\bibfnamefont {D.~A.}\ \bibnamefont {Chenet}},
  \bibinfo {author} {\bibfnamefont {A.~M.}\ \bibnamefont {van~der Zande}},
  \bibinfo {author} {\bibfnamefont {J.~C.}\ \bibnamefont {Hone}}, \ and\
  \bibinfo {author} {\bibfnamefont {T.~F.}\ \bibnamefont {Heinz}},\ }\href
  {\doibase 10.1021/acsphotonics.5b00486} {\bibfield  {journal} {\bibinfo
  {journal} {ACS Photonics}\ }\textbf {\bibinfo {volume} {3}},\ \bibinfo
  {pages} {96} (\bibinfo {year} {2016})}\BibitemShut {NoStop}%
\bibitem [{\citenamefont {Zhao}\ \emph {et~al.}(2015)\citenamefont {Zhao},
  \citenamefont {Wu}, \citenamefont {Zhong}, \citenamefont {Guo}, \citenamefont
  {Wang}, \citenamefont {Xia}, \citenamefont {Yang}, \citenamefont {Tan},\ and\
  \citenamefont {Wang}}]{Zhao2015}%
  \BibitemOpen
  \bibfield  {author} {\bibinfo {author} {\bibfnamefont {H.}~\bibnamefont
  {Zhao}}, \bibinfo {author} {\bibfnamefont {J.}~\bibnamefont {Wu}}, \bibinfo
  {author} {\bibfnamefont {H.}~\bibnamefont {Zhong}}, \bibinfo {author}
  {\bibfnamefont {Q.}~\bibnamefont {Guo}}, \bibinfo {author} {\bibfnamefont
  {X.}~\bibnamefont {Wang}}, \bibinfo {author} {\bibfnamefont {F.}~\bibnamefont
  {Xia}}, \bibinfo {author} {\bibfnamefont {L.}~\bibnamefont {Yang}}, \bibinfo
  {author} {\bibfnamefont {P.}~\bibnamefont {Tan}}, \ and\ \bibinfo {author}
  {\bibfnamefont {H.}~\bibnamefont {Wang}},\ }\href {\doibase
  10.1007/s12274-015-0865-0} {\bibfield  {journal} {\bibinfo  {journal} {Nano
  Research}\ }\textbf {\bibinfo {volume} {8}},\ \bibinfo {pages} {3651}
  (\bibinfo {year} {2015})}\BibitemShut {NoStop}%
\bibitem [{\citenamefont {Nagler}\ \emph {et~al.}(2016)\citenamefont {Nagler},
  \citenamefont {Plechinger}, \citenamefont {Schüller},\ and\ \citenamefont
  {Korn}}]{PSSR:PSSR201510412}%
  \BibitemOpen
  \bibfield  {author} {\bibinfo {author} {\bibfnamefont {P.}~\bibnamefont
  {Nagler}}, \bibinfo {author} {\bibfnamefont {G.}~\bibnamefont {Plechinger}},
  \bibinfo {author} {\bibfnamefont {C.}~\bibnamefont {Schüller}}, \ and\
  \bibinfo {author} {\bibfnamefont {T.}~\bibnamefont {Korn}},\ }\href {\doibase
  10.1002/pssr.201510412} {\bibfield  {journal} {\bibinfo  {journal} {Phys.
  Status Solidi Rapid Res. Lett.}\ }\textbf {\bibinfo {volume} {10}},\ \bibinfo
  {pages} {185} (\bibinfo {year} {2016})}\BibitemShut {NoStop}%
\bibitem [{\citenamefont {He}\ \emph {et~al.}(2016)\citenamefont {He},
  \citenamefont {Yan}, \citenamefont {Yin}, \citenamefont {Ye}, \citenamefont
  {Ye}, \citenamefont {Cheng}, \citenamefont {Li},\ and\ \citenamefont
  {Lui}}]{doi:10.1021/acs.nanolett.5b04925}%
  \BibitemOpen
  \bibfield  {author} {\bibinfo {author} {\bibfnamefont {R.}~\bibnamefont
  {He}}, \bibinfo {author} {\bibfnamefont {J.-A.}\ \bibnamefont {Yan}},
  \bibinfo {author} {\bibfnamefont {Z.}~\bibnamefont {Yin}}, \bibinfo {author}
  {\bibfnamefont {Z.}~\bibnamefont {Ye}}, \bibinfo {author} {\bibfnamefont
  {G.}~\bibnamefont {Ye}}, \bibinfo {author} {\bibfnamefont {J.}~\bibnamefont
  {Cheng}}, \bibinfo {author} {\bibfnamefont {J.}~\bibnamefont {Li}}, \ and\
  \bibinfo {author} {\bibfnamefont {C.~H.}\ \bibnamefont {Lui}},\ }\href
  {\doibase 10.1021/acs.nanolett.5b04925} {\bibfield  {journal} {\bibinfo
  {journal} {Nano Letters}\ }\textbf {\bibinfo {volume} {16}},\ \bibinfo
  {pages} {1404} (\bibinfo {year} {2016})},\ \Eprint
  {http://arxiv.org/abs/http://dx.doi.org/10.1021/acs.nanolett.5b04925}
  {http://dx.doi.org/10.1021/acs.nanolett.5b04925} \BibitemShut {NoStop}%
\bibitem [{\citenamefont {Lorchat}\ \emph {et~al.}(2016)\citenamefont
  {Lorchat}, \citenamefont {Froehlicher},\ and\ \citenamefont
  {Berciaud}}]{doi:10.1021/acsnano.5b07844}%
  \BibitemOpen
  \bibfield  {author} {\bibinfo {author} {\bibfnamefont {E.}~\bibnamefont
  {Lorchat}}, \bibinfo {author} {\bibfnamefont {G.}~\bibnamefont
  {Froehlicher}}, \ and\ \bibinfo {author} {\bibfnamefont {S.}~\bibnamefont
  {Berciaud}},\ }\href {\doibase 10.1021/acsnano.5b07844} {\bibfield  {journal}
  {\bibinfo  {journal} {ACS Nano}\ }\textbf {\bibinfo {volume} {10}},\ \bibinfo
  {pages} {2752} (\bibinfo {year} {2016})},\ \Eprint
  {http://arxiv.org/abs/http://dx.doi.org/10.1021/acsnano.5b07844}
  {http://dx.doi.org/10.1021/acsnano.5b07844} \BibitemShut {NoStop}%
\bibitem [{\citenamefont {Lin}\ \emph {et~al.}(2015)\citenamefont {Lin},
  \citenamefont {Komsa}, \citenamefont {Yeh}, \citenamefont {Björkman},
  \citenamefont {Liang}, \citenamefont {Ho}, \citenamefont {Huang},
  \citenamefont {Chiu}, \citenamefont {Krasheninnikov},\ and\ \citenamefont
  {Suenaga}}]{doi:10.1021/acsnano.5b04851}%
  \BibitemOpen
  \bibfield  {author} {\bibinfo {author} {\bibfnamefont {Y.-C.}\ \bibnamefont
  {Lin}}, \bibinfo {author} {\bibfnamefont {H.-P.}\ \bibnamefont {Komsa}},
  \bibinfo {author} {\bibfnamefont {C.-H.}\ \bibnamefont {Yeh}}, \bibinfo
  {author} {\bibfnamefont {T.}~\bibnamefont {Björkman}}, \bibinfo {author}
  {\bibfnamefont {Z.-Y.}\ \bibnamefont {Liang}}, \bibinfo {author}
  {\bibfnamefont {C.-H.}\ \bibnamefont {Ho}}, \bibinfo {author} {\bibfnamefont
  {Y.-S.}\ \bibnamefont {Huang}}, \bibinfo {author} {\bibfnamefont {P.-W.}\
  \bibnamefont {Chiu}}, \bibinfo {author} {\bibfnamefont {A.~V.}\ \bibnamefont
  {Krasheninnikov}}, \ and\ \bibinfo {author} {\bibfnamefont {K.}~\bibnamefont
  {Suenaga}},\ }\href {\doibase 10.1021/acsnano.5b04851} {\bibfield  {journal}
  {\bibinfo  {journal} {ACS Nano}\ }\textbf {\bibinfo {volume} {9}},\ \bibinfo
  {pages} {11249} (\bibinfo {year} {2015})},\ \Eprint
  {http://arxiv.org/abs/http://dx.doi.org/10.1021/acsnano.5b04851}
  {http://dx.doi.org/10.1021/acsnano.5b04851} \BibitemShut {NoStop}%
\bibitem [{\citenamefont {Tongay}\ \emph {et~al.}(2014)\citenamefont {Tongay},
  \citenamefont {Sahin}, \citenamefont {Ko}, \citenamefont {Luce},
  \citenamefont {Fan}, \citenamefont {Liu}, \citenamefont {Zhou}, \citenamefont
  {Huang}, \citenamefont {Ho}, \citenamefont {Yan}, \citenamefont {Ogletree},
  \citenamefont {Aloni}, \citenamefont {Ji}, \citenamefont {Li}, \citenamefont
  {Li}, \citenamefont {Peeters},\ and\ \citenamefont
  {Wu}}]{doi:10.1038/ncomms4252}%
  \BibitemOpen
  \bibfield  {author} {\bibinfo {author} {\bibfnamefont {S.}~\bibnamefont
  {Tongay}}, \bibinfo {author} {\bibfnamefont {H.}~\bibnamefont {Sahin}},
  \bibinfo {author} {\bibfnamefont {C.}~\bibnamefont {Ko}}, \bibinfo {author}
  {\bibfnamefont {A.}~\bibnamefont {Luce}}, \bibinfo {author} {\bibfnamefont
  {W.}~\bibnamefont {Fan}}, \bibinfo {author} {\bibfnamefont {K.}~\bibnamefont
  {Liu}}, \bibinfo {author} {\bibfnamefont {J.}~\bibnamefont {Zhou}}, \bibinfo
  {author} {\bibfnamefont {Y.-S.}\ \bibnamefont {Huang}}, \bibinfo {author}
  {\bibfnamefont {C.-H.}\ \bibnamefont {Ho}}, \bibinfo {author} {\bibfnamefont
  {J.}~\bibnamefont {Yan}}, \bibinfo {author} {\bibfnamefont {D.~F.}\
  \bibnamefont {Ogletree}}, \bibinfo {author} {\bibfnamefont {S.}~\bibnamefont
  {Aloni}}, \bibinfo {author} {\bibfnamefont {J.}~\bibnamefont {Ji}}, \bibinfo
  {author} {\bibfnamefont {S.}~\bibnamefont {Li}}, \bibinfo {author}
  {\bibfnamefont {J.}~\bibnamefont {Li}}, \bibinfo {author} {\bibfnamefont
  {F.~M.}\ \bibnamefont {Peeters}}, \ and\ \bibinfo {author} {\bibfnamefont
  {J.}~\bibnamefont {Wu}},\ }\href {\doibase 10.1038/ncomms4252} {\bibfield
  {journal} {\bibinfo  {journal} {Nat. Commun.}\ }\textbf {\bibinfo {volume}
  {5}} (\bibinfo {year} {2014}),\ 10.1038/ncomms4252}\BibitemShut {NoStop}%
\bibitem [{\citenamefont {Gutiérrez-Lezama}\ \emph {et~al.}(2016)\citenamefont
  {Gutiérrez-Lezama}, \citenamefont {Reddy}, \citenamefont {Ubrig},\ and\
  \citenamefont {Morpurgo}}]{2053-1583-3-4-045016}%
  \BibitemOpen
  \bibfield  {author} {\bibinfo {author} {\bibfnamefont {I.}~\bibnamefont
  {Gutiérrez-Lezama}}, \bibinfo {author} {\bibfnamefont {B.~A.}\ \bibnamefont
  {Reddy}}, \bibinfo {author} {\bibfnamefont {N.}~\bibnamefont {Ubrig}}, \ and\
  \bibinfo {author} {\bibfnamefont {A.~F.}\ \bibnamefont {Morpurgo}},\ }\href
  {http://stacks.iop.org/2053-1583/3/i=4/a=045016} {\bibfield  {journal}
  {\bibinfo  {journal} {2D Materials}\ }\textbf {\bibinfo {volume} {3}},\
  \bibinfo {pages} {045016} (\bibinfo {year} {2016})}\BibitemShut {NoStop}%
\bibitem [{\citenamefont {Qiao}\ \emph {et~al.}(2016)\citenamefont {Qiao},
  \citenamefont {Wu}, \citenamefont {Zhou}, \citenamefont {Qiao}, \citenamefont
  {Shi}, \citenamefont {Chen}, \citenamefont {Zhang}, \citenamefont {Zhang},
  \citenamefont {Ji},\ and\ \citenamefont {Tan}}]{C6NR01569G}%
  \BibitemOpen
  \bibfield  {author} {\bibinfo {author} {\bibfnamefont {X.-F.}\ \bibnamefont
  {Qiao}}, \bibinfo {author} {\bibfnamefont {J.-B.}\ \bibnamefont {Wu}},
  \bibinfo {author} {\bibfnamefont {L.}~\bibnamefont {Zhou}}, \bibinfo {author}
  {\bibfnamefont {J.}~\bibnamefont {Qiao}}, \bibinfo {author} {\bibfnamefont
  {W.}~\bibnamefont {Shi}}, \bibinfo {author} {\bibfnamefont {T.}~\bibnamefont
  {Chen}}, \bibinfo {author} {\bibfnamefont {X.}~\bibnamefont {Zhang}},
  \bibinfo {author} {\bibfnamefont {J.}~\bibnamefont {Zhang}}, \bibinfo
  {author} {\bibfnamefont {W.}~\bibnamefont {Ji}}, \ and\ \bibinfo {author}
  {\bibfnamefont {P.-H.}\ \bibnamefont {Tan}},\ }\href {\doibase
  10.1039/C6NR01569G} {\bibfield  {journal} {\bibinfo  {journal} {Nanoscale}\
  }\textbf {\bibinfo {volume} {8}},\ \bibinfo {pages} {8324} (\bibinfo {year}
  {2016})}\BibitemShut {NoStop}%
\bibitem [{\citenamefont {Ovchinnikov}\ \emph {et~al.}(2016)\citenamefont
  {Ovchinnikov}, \citenamefont {Gargiulo}, \citenamefont {Allain},
  \citenamefont {Pasquier}, \citenamefont {Dumcenco}, \citenamefont {Ho},
  \citenamefont {Yazyev},\ and\ \citenamefont {Kis}}]{Ovchinnikov2016}%
  \BibitemOpen
  \bibfield  {author} {\bibinfo {author} {\bibfnamefont {D.}~\bibnamefont
  {Ovchinnikov}}, \bibinfo {author} {\bibfnamefont {F.}~\bibnamefont
  {Gargiulo}}, \bibinfo {author} {\bibfnamefont {A.}~\bibnamefont {Allain}},
  \bibinfo {author} {\bibfnamefont {D.~J.}\ \bibnamefont {Pasquier}}, \bibinfo
  {author} {\bibfnamefont {D.}~\bibnamefont {Dumcenco}}, \bibinfo {author}
  {\bibfnamefont {C.-H.}\ \bibnamefont {Ho}}, \bibinfo {author} {\bibfnamefont
  {O.~V.}\ \bibnamefont {Yazyev}}, \ and\ \bibinfo {author} {\bibfnamefont
  {A.}~\bibnamefont {Kis}},\ }\href {http://dx.doi.org/10.1038/ncomms12391}
  {\bibfield  {journal} {\bibinfo  {journal} {Nat. Commun.}\ }\textbf {\bibinfo
  {volume} {7}},\ \bibinfo {pages} {12391} (\bibinfo {year}
  {2016})}\BibitemShut {NoStop}%
\bibitem [{\citenamefont {Tusche}\ \emph {et~al.}(2015)\citenamefont {Tusche},
  \citenamefont {Krasyuk},\ and\ \citenamefont {Kirschner}}]{Tusche2015520}%
  \BibitemOpen
  \bibfield  {author} {\bibinfo {author} {\bibfnamefont {C.}~\bibnamefont
  {Tusche}}, \bibinfo {author} {\bibfnamefont {A.}~\bibnamefont {Krasyuk}}, \
  and\ \bibinfo {author} {\bibfnamefont {J.}~\bibnamefont {Kirschner}},\ }\href
  {\doibase http://dx.doi.org/10.1016/j.ultramic.2015.03.020} {\bibfield
  {journal} {\bibinfo  {journal} {Ultramicroscopy}\ }\textbf {\bibinfo {volume}
  {159, Part 3}},\ \bibinfo {pages} {520 } (\bibinfo {year}
  {2015})}\BibitemShut {NoStop}%
\bibitem [{\citenamefont {Seah}\ and\ \citenamefont
  {Dench}(1979)}]{SIA:SIA740010103}%
  \BibitemOpen
  \bibfield  {author} {\bibinfo {author} {\bibfnamefont {M.~P.}\ \bibnamefont
  {Seah}}\ and\ \bibinfo {author} {\bibfnamefont {W.~A.}\ \bibnamefont
  {Dench}},\ }\href {\doibase 10.1002/sia.740010103} {\bibfield  {journal}
  {\bibinfo  {journal} {Surface and Interface Analysis}\ }\textbf {\bibinfo
  {volume} {1}},\ \bibinfo {pages} {2} (\bibinfo {year} {1979})}\BibitemShut
  {NoStop}%
\bibitem [{\citenamefont {Castellanos-Gomez}\ \emph {et~al.}(2014)\citenamefont
  {Castellanos-Gomez}, \citenamefont {Buscema}, \citenamefont {Molenaar},
  \citenamefont {Singh}, \citenamefont {Janssen}, \citenamefont {van~der
  Zant},\ and\ \citenamefont {Steele}}]{2053-1583-1-1-011002}%
  \BibitemOpen
  \bibfield  {author} {\bibinfo {author} {\bibfnamefont {A.}~\bibnamefont
  {Castellanos-Gomez}}, \bibinfo {author} {\bibfnamefont {M.}~\bibnamefont
  {Buscema}}, \bibinfo {author} {\bibfnamefont {R.}~\bibnamefont {Molenaar}},
  \bibinfo {author} {\bibfnamefont {V.}~\bibnamefont {Singh}}, \bibinfo
  {author} {\bibfnamefont {L.}~\bibnamefont {Janssen}}, \bibinfo {author}
  {\bibfnamefont {H.~S.~J.}\ \bibnamefont {van~der Zant}}, \ and\ \bibinfo
  {author} {\bibfnamefont {G.~A.}\ \bibnamefont {Steele}},\ }\href
  {http://stacks.iop.org/2053-1583/1/i=1/a=011002} {\bibfield  {journal}
  {\bibinfo  {journal} {2D Materials}\ }\textbf {\bibinfo {volume} {1}},\
  \bibinfo {pages} {011002} (\bibinfo {year} {2014})}\BibitemShut {NoStop}%
\bibitem [{FLE()}]{FLEURhomepage}%
  \BibitemOpen
  \href@noop {} {\enquote {\bibinfo {title} {{FLEUR project homepage}},}\
  }\bibinfo {howpublished} {\url{http://www.flapw.de/}},\ \bibinfo {note}
  {accessed: 2016-10-03}\BibitemShut {NoStop}%
\bibitem [{\citenamefont {Friedrich}\ \emph {et~al.}(2010)\citenamefont
  {Friedrich}, \citenamefont {Bl\"ugel},\ and\ \citenamefont
  {Schindlmayr}}]{friedrich2010-spex}%
  \BibitemOpen
  \bibfield  {author} {\bibinfo {author} {\bibfnamefont {C.}~\bibnamefont
  {Friedrich}}, \bibinfo {author} {\bibfnamefont {S.}~\bibnamefont {Bl\"ugel}},
  \ and\ \bibinfo {author} {\bibfnamefont {A.}~\bibnamefont {Schindlmayr}},\
  }\href@noop {} {\bibfield  {journal} {\bibinfo  {journal} {Phys. Rev. B}\
  }\textbf {\bibinfo {volume} {81}},\ \bibinfo {pages} {125102} (\bibinfo
  {year} {2010})}\BibitemShut {NoStop}%
\bibitem [{\citenamefont {Mostofi}\ \emph {et~al.}(2008)\citenamefont
  {Mostofi}, \citenamefont {Yates}, \citenamefont {Lee}, \citenamefont {Souza},
  \citenamefont {Vanderbilt},\ and\ \citenamefont {Marzari}}]{wannier90}%
  \BibitemOpen
  \bibfield  {author} {\bibinfo {author} {\bibfnamefont {A.~A.}\ \bibnamefont
  {Mostofi}}, \bibinfo {author} {\bibfnamefont {J.~R.}\ \bibnamefont {Yates}},
  \bibinfo {author} {\bibfnamefont {Y.-S.}\ \bibnamefont {Lee}}, \bibinfo
  {author} {\bibfnamefont {I.}~\bibnamefont {Souza}}, \bibinfo {author}
  {\bibfnamefont {D.}~\bibnamefont {Vanderbilt}}, \ and\ \bibinfo {author}
  {\bibfnamefont {N.}~\bibnamefont {Marzari}},\ }\href@noop {} {\bibfield
  {journal} {\bibinfo  {journal} {Comput. Phys. Commun.}\ }\textbf {\bibinfo
  {volume} {178}},\ \bibinfo {pages} {685} (\bibinfo {year}
  {2008})}\BibitemShut {NoStop}%
\end{thebibliography}%

\end{document}